\documentclass[preprint,eqsecnum,preprintnumbers,nofootinbib,byrevtex,prd,aps,showpacs,showkeys,groupedaddress,floatfix]{revtex4}
\usepackage{bm}
\usepackage{graphics}
\usepackage{graphicx}
\usepackage{epsfig}
\usepackage{amssymb}
\usepackage{amsmath}
\begin{document}

\date{}
\title{Analytical solutions of the  Schr\"{o}dinger
equation  with the Manning-Rosen  potential plus a Ring-Shaped like
potential}
\author{H.~I.~Ahmadov$^{1}$}\email{E-mail:hikmatahmadov@yahoo.com}
\author{C.~Aydin$^{2,3}$} \email{E-mail:coskun@ktu.edu.tr}
\author{N.~Sh.~Huseynova$^{4}$}\email{E-mail:nargiz_huseynova@yahoo.com}
\author{O.~Uzun$^{2}$}\email{E-mail:oguzhan_deu@hotmail.com}

\affiliation{$^{1}$Department  of Equations of Mathematical Physics,
Faculty of Applied Mathematics and Cybernetics, Baku State
University, Z. Khalilov st. 23, AZ-1148, Baku, Azerbaijan}
\affiliation{$^{2}$ Department of Physics, Karadeniz Technical
University, 61080, Trabzon, Turkey and Department of Physics}
\affiliation{$^{3}$ University of Surrey, Guildford Surrey GU2 7XH,
United Kingdom} \affiliation{$^{4}$Institute  of Applied Mathematics
Baku State University, Z. Khalilov st. 23, AZ-1148, Baku,
Azerbaijan}


\begin{abstract}
The analytical solution of the Schr\"{o}dinger equation for the
Manning-Rosen potential plus a ring-shaped like potential is
obtained by applying the Nikiforov-Uvarov method by using the
improved approximation scheme to the centrifugal potential for
arbitrary $l$ states. The energy levels are worked out and the
corresponding normalized eigenfunctions are obtained in terms of
orthogonal polynomials for arbitrary $l$ states.
\end{abstract}


\maketitle

\section{\bf Introduction}

As known, one of the main objectives in theoretical physics since
the early years of quantum mechanics (QM) is to obtain an exact
solution of the Schr\"{o}dinger equation (SE) for some special
potentials of physical interest. Since the wave function contains
all necessary information for full description of a quantum system,
an analytical solution of the SE is of high importance in
non-relativistic and relativistic quantum
mechanics~\cite{Greiner,Bagrov}.  There are few potentials for which
the SE can be solved explicitly for all $n$ and $l$ quantum states.

The non-central potentials are needed to obtain better results than
central potentials about the dynamical properties of the molecular
structures and interactions. Researchers added ring shaped
potentials to certain potentials, i.e, Coulomb,
Kratzer~\cite{Flugge} and Manning-Rosen potentials~\cite{Manning} to
obtain non-central potentials. Ring-shaped potentials can be used in
quantum chemistry to describe the ring shaped organic molecules such
as benzene and in nuclear physics to investigate the interaction
between deformed pair of nucleus and spin orbit coupling for the
motion of the particle in the potential fields.

It would be interesting and important to solve the SE for the
Manning-Rosen potential plus a Ring-Shaped like potential for
$l\neq0$, since it has been extensively used to describe the bound
and continuum states of the interacting systems. Thus, one can
obtain the energy eigenvalues and corresponding eigenfunctions of
the one particle problem within this potential. The central
Manning-Rosen  potential is defined by
\begin{equation}
V(r)=\frac{1}{kb^2}\left[\frac{\alpha(\alpha-1)exp(-2r/b)}{(1-exp(-r/b))^2}-\frac{Aexp(-r/b)}{(1-exp(-r/b)}\right],
k=2\mu/\hbar^2.
\end{equation}
where $A$ and $\alpha$ are dimensionless parameters, and $b$ is the
screening parameter. This potential is used as a mathematical
modeling  of the diatomic molecular vibrations and it constitutes a
convenient model for other physical situations. It is known that,
using for this potential the SE can be solved exactly for s-wave ($l
= 0$) ~\cite{Dong1}. Unfortunately, for an arbitrary $l$-states
($l\neq 0$), the SE does not admit an exact solution. In such a
case, the SE can be solved numerically or approximately using
approximation schemes~\cite{Dong2}.

The potential which is solved in this study is obtained by adding
ring-shaped potential term ~\cite{Makarov} as,

\begin{equation}
V(r,\theta)=\frac{1}{k}\left[\frac{\alpha(\alpha-1)exp(-2r/b)}{b^2
(1-exp(-r/b))^2}-\frac{Aexp(-r/b)}{b^2
(1-exp(-r/b)}+\frac{\beta'}{r^2sin^2\theta}+\frac{\beta
cos\theta}{r^2sin^2\theta}\right].
\end{equation}

So far, many methods were developed and applied, such as
supersymmetry (SUSY)~\cite{Cooper,Morales},
factorization~\cite{Dong3} , Laplace transform approach~\cite{Arda}
and the path integral method~\cite{Cai}, to solve the radial SE
exactly or quasi-exactly for $l\neq0$ within these potentials. An
other method known as the Nikiforov-Uvarov (NU) method
~\cite{Nikiforov} was proposed for solving the SE analytically. Many
works show the power and simplicity of NU method in solving central
and noncentral potentials~\cite{Badalov1,Badalov2,Badalov3}. This
method is based on solving the second order linear differential
equation by reducing to a generalized equation of hypergeometric
type which is a second order homogeneous differential equation with
polynomials coefficients of degree not exceeding the corresponding
order of differentiation.

In this study, we obtain the energy eigenvalues and corresponding
eigenfunctions for arbitrary $l$ states by solving the SE for the
Manning-Rosen potential plus a ring-shaped like potential using NU
method. Moreover, by chancing parameters we also obtain solutions
for Hulth\'en potential~\cite{Varshni}, and Hulth\'en  plus ring
shaped potential. It should be noted, that same problem have been
studied in Ref.~\cite{Antia} as well, but our results disagree with
those conducted in Ref.~\cite{Antia}

 The organization of this paper is as follows. The SE within Manning
Rosen plus a ring-shaped like potential is provided in Section
\ref{ht}. Bound state solution of the radial SE by NU method is
presented in Section \ref{ir}. The solution of angle-dependent part
of the SE is presented in Section \ref{ar} and, the numerical
results for energy levels and the corresponding normalized
eigenfunctions are presented in Section \ref{br}. Finally, some
concluding remarks are stated in Section \ref{dr}.

\section{\bf The Schr\"{o}dinger equation with the Manning-Rosen potential
plus a ring-shaped like potential}\label{ht}

The Schr\"{o}dinger equation in spherical coordinates is given as

\begin{equation}
  \bigtriangledown ^2 \psi+  \frac{2\mu}{\hbar^2}[E-V(r,\theta
  )]\psi=0.
\end{equation}
Considering this equation, the total wave function is written as
\begin{equation}
  \psi (r,\theta ,\phi )=R(r)\Theta (\theta )\Phi (\phi ),
\end{equation}
where the polar angle solution is given by
\begin{equation}
  \Phi (\phi )=\frac{1}{2\pi }e^{im\phi }, m=0,\pm 1,\pm 2,...
\end{equation}
Thus for radial and azimuthal  SE for Manning Rosen plus ring-shaped
like potential are
\begin{equation}
  R''(r)+\frac{2}{r}R'(r)+\left[\frac{2\mu }{\hbar ^2}E+ \frac{1}{b^{2}}\frac{Ae^{-r/b}}{1-e^{-r/b}}-\frac{1}{b^2}\frac{\alpha(\alpha-1)e^{-2r/b}}{(1-e^{-r/b})^2}
  -\frac{\lambda}{r^2}\right]R(r)=0,
\end{equation}
\begin{equation}
  \Theta''(\theta)+cot\theta \Theta '(\theta)+\left[-\left(\frac{\beta^\prime+\beta cos\theta}{sin^2\theta}\right)+\lambda-\frac{m^2}{sin^2\theta}\right]\Theta
  (\theta)=0,
\end{equation}

respectively.

\section{\bf Bound state Solution of the Radial Schr\"{o}dinger
equation.}\label{ir}

As know, Eq.(2.4) is the radial SE for Manning-Rosen plus a
ring-shaped potential. In order to solve Eq.(2.4) with
$\lambda=l(l+1)\neq0,$ we must make an approximation for the
centrifugal term. When $r/b < < 1$, we use an improved approximation
scheme~\cite{Jia} to deal with the centrifugal term
\begin{equation}
\left[C_0+\frac{{e^{ - r/b} }}{{(1 - e^{ - r/b} )^2 }} \right]
\approx \frac{{b^2 }}{{r^2 }} + \left( {C_0 - \frac{1}{{12}}}
\right) + O(\frac{r^2}{b^2} ),\,\,C_0 = \frac{1}{{12}},\frac{1}{{r^2
}} \approx \frac{1}{{b^2 }}\left[ {C_o + \frac{{e^{ - r/b} }}{{(1 -
e^{ - r/b} )^2 }}} \right],
\end{equation}

where the parameter $C_0=\frac{1}{12}$ is a dimensionless constant.
However, when $C_0=0$ the approximation scheme becomes
the convectional approximation scheme suggested by Greene and
Aldrich~\cite{Greene}.

We assume $R(r)=\frac{1}{r}\chi(r)$ in
Eq.(2.4) and the radial SE  becomes
\begin{equation}
\chi ''(r) +\left[ {\frac{2\mu }{\hbar ^2}E + \frac{1}{{b^2}}\frac{{Ae^{ - r/b} }}{{(1 -
e^{r/b} )}} - \frac{1}{{b^2 }}\frac{{\alpha (\alpha  - 1)e^{-2r/b}
}}{{(1 - e^{ - r/b} )^2 }} - \frac{\lambda
}{{b{}^2}}}\left[ {C_o + \frac{{e^{ - r/b} }}{{(1 - e^{ - r/b} )^2
}}} \right] \right]\chi (r) = 0.
\end{equation}

In order to transform Eq.(3.2) in to the equation of the generalized
hypergeometric-type which is in the form ~\cite{Nikiforov}
\begin{equation}
  \chi''(s)+\frac{\tilde{\tau}}{\sigma } \chi'(s)+\frac{\tilde{\sigma}}{\sigma ^2} \chi(s)=0,
\end{equation} we use the transformation $ s = e^{- r/b}$. Hence we obtain

\begin{eqnarray}
\chi''(s)+\chi
'(s)\frac{1-s}{s(1-s)}+\biggl[\frac{1}{s(1-s)}\biggr]^2\biggl[-\epsilon^2(1-s)^2+
As(1-s)-\alpha (\alpha -1)s^2-  \nonumber \\
(1-s)^2\lambda \biggl(C_0+\frac{s}{(1-s)^2}\biggr)\biggr]\chi(s)=0,
\end{eqnarray}
where we use the following notation for bound states
\begin{equation}
-\epsilon ^2=\frac{2\mu }{\hbar ^2}Eb^2,\,\,\, E<0.
\end{equation}
Now, we can successfully apply NU method of definition for
eigenvalues of energy. By comparing Eq.(3.4) with Eq.(3.3) we can
define the following:

$\tilde{\tau} (s) = 1 - s,\sigma (s) = s(1 - s) $,
\begin{equation}
   \tilde{ \sigma} (s) = s^2 [ - \epsilon ^2  -A -\alpha (\alpha -1) - \lambda c_0 ] + s[2\epsilon ^2  + A + 2\lambda C_0  - \lambda ] + [-\epsilon ^2  - \lambda c_0 ] .
\end{equation}

If we take the following factorization \begin{equation}
  \chi (s)=\phi (s)y(s),
\end{equation}
for the appropriate function $\phi (s)$ the Eq.(3.4) takes the form
of the well known hypergeometric-type equation. The appropriate
$\phi (s)$ function must satisfy the following condition:
\begin{equation}
 \frac { \phi ^{'} (s)}{\phi (s)}=\frac {\pi (s)}{\sigma (s)},
\end{equation}
where function $\pi (s)$ is defined as
\begin{equation}
  \pi(s)= \frac{{ \sigma' -\tilde{\tau }}}{2} \pm \sqrt {  (\frac{{ \sigma' -\tilde{\tau} }}{2} )^2 -\tilde{\sigma} +k\sigma }.
\end{equation}
Finally the equation, where $y(s)$ is one of its solutions, takes
the form known as hypergeometric-type,
\begin{equation}
  \sigma (s) y^{''} (s) + \tau (s) y^{'} (s) +\bar{\lambda} y(s)=0,
\end{equation}
where
\begin{equation}
  \bar{\lambda} =k+\pi ^{'}
\end{equation}
and
\begin{equation}
  \tau (s)=\tilde{\tau} (s) +2\pi (s).
\end{equation}

For our problem, the $\pi (s)$ function is written as
\begin{equation}
  \pi(s)= \frac{{ - s}}{2} \pm \sqrt {s^2 [a - k] - s[b -k] + c},
\end{equation}

where the values of the parameters are
$$
 a = \frac{1}{4} +{\epsilon ^2 + A+ \alpha( \alpha-1)+ \lambda C_0},
$$
$$
 b = 2\epsilon ^2 + A+ 2\lambda C_0 -\lambda,
$$
$$
 c = \epsilon ^2  + \lambda C_0.
$$

 The constant parameter $k$ can be found complying with the
condition that the discriminant of the expression under the square
root is equal to zero. Hence, we obtain

\begin{equation}
k_{1,2}  = (b - 2c) \pm 2\sqrt {c^2  + c(a - b)}.
\end{equation}

Now, we can find four possible functions for $\pi(s)$:

\begin{equation}
\pi (s) = \frac{{ - s}}{2} \pm \left\{ \begin{array}{l}
 (\sqrt c  - \sqrt {c + a - b} )s - \sqrt c \,\,\,for\,\,\, k = (b - 2c) + 2\sqrt {c^2  + c(a - b)} , \\
 (\sqrt c  + \sqrt {c + a - b} )s - \sqrt c \,\,\, for\,\,\, k = (b - 2c) - 2\sqrt {c^2  + c(a - b)} . \\
 \end{array} \right.
\end{equation}

According to NU method, from the four possible forms of the
polynomial $\pi(s)$, we select the one for which the function $\tau
(s)$  has the negative derivative. Therefore, the appropriate
function $\pi(s)$ and $\tau(s)$ are

\begin{equation}
  \pi(s)=\sqrt{c}-s\left[\frac{1}{2}+\sqrt{c}+\sqrt{c+a-b}\right],
\end{equation}

\begin{equation}
\tau (s) = 1 +2\sqrt c -2s \left[ 1+\sqrt{c+a-b}\right]  ,
\end{equation}

for
\begin{equation}
k  = (b - 2c) -2\sqrt {c^2  + c(a - b)}.
\end{equation}

Also by Eq.(3.11) we can define the constant $ \bar{\lambda} $ as

\begin{equation}
 \bar{ \lambda}=b-2c- 2\sqrt {c^2  + c(a - b)}  - \left[\frac{1}{2} + {\sqrt{c} + \sqrt {c + a - b} }\right].
\end{equation}

Given a nonnegative integer $n$, the hypergeometric-type equation has a unique polynomials solution of degree $n$ if and only if

\begin{equation}
  \bar{\lambda}=\bar{\lambda}_n=-n\tau'-\frac{n(n-1)}{2}\sigma '', (n=0,1,2...)
\end{equation}

and $\bar{\lambda}_m\neq\bar{\lambda}_n $ for
~$m=0,1,2,...,n-1$~\cite{Area}, then it follows that
$$
\bar{\lambda} _{n_{r}}  = b-2c- 2\sqrt {c^2  + c(a - b)}
-\left[\frac{1}{2} + {\sqrt{c} + \sqrt {c + a - b} }\right]
$$
\begin{equation}
= 2n_r\left[ {1 +\left( {\sqrt c  + \sqrt {c + a - b} } \right)}
\right] + n_r(n_r - 1).
\end{equation}

We can solve Eq.(3.21) explicitly for $c$ and by using the relation
 $c=\epsilon ^2  + \lambda C_0$, which brings

\begin{equation}
  \epsilon^{2}  = \left[ \frac{\lambda+1/2+\Lambda  (1 + 2n_r) + n_r(n_r + 1)-A}{ 2\Lambda + 1 + 2n} \right]^2  -\lambda
  C_0,
\end{equation}
where $\Lambda =\sqrt{1/4+\alpha (\alpha -1)+\lambda }$.

We substitute $\epsilon ^2$ into Eq.(3.5) with $\lambda =l(l+1)$,
which identifies

\begin{equation}
E_{n_{r},l} = \frac{-h^2 }{2\mu b^2}\left[\left[
 n_{r}+1/2+\frac{(l-n_{r})(l+n_{r}+1)-A}{ 2\Lambda + 1 + 2n_{r}} \right]^2
  -l(l+1) C_0\right].
\end{equation}

Now, using NU method we can obtain the radial eigenfunctions. After
substituting   $\pi(s)$ and $\sigma(s) $ into Eq.(3.8) and solving
first order differential equation, it is easy to obtain

\begin{equation}
  \phi (s)=s^{\sqrt c}(1-s)^K,
\end{equation}

where $K=1/2+\Lambda $.

 Furthermore, the other part of the wave function y(s) is the hypergeometric-type function whose polynomial solutions are given by Rodrigues relation
 \begin{equation}
   y_{n}(s) = \frac {B_{n}}{\rho (s)} \frac{{d^{n} }}{{ds^{n}
}}\left[ \sigma ^{n}(s)\rho (s) \right],
 \end{equation}
 where $B_n$ is a normalizing constant and $\rho (s)$ is the weight function which is the solutions of the Pearson
 differential equation. The Pearson
 differential equation and $\rho(s)$ for our problem is given as

 \begin{equation}
   (\sigma \rho )^{'} =\tau \rho ,
 \end{equation}

 \begin{equation}
  \rho(s) =(1 -s)^{2K - 1} s^{2\sqrt c },
\end{equation}
respectively.

Substituting Eq.(3.27) in Eq.(3.25) we get

\begin{equation}
y_{n_{r}}(s) = B_{n_{r}}(1 - s)^{1 - 2K} s^{2\sqrt c }
\frac{{d^{n_{r}} }}{{ds^{n_{r}} }}\left[ {s^{2\sqrt c  + n_{r}} (1 -
s)^{2K - 1 + n_{r}} } \right].
\end{equation}
Then by using the following definition of the Jacobi  polynomials
~\cite{Abramowitz}:

\begin{equation}
P_n^{(a,b)} (s) = \frac{( - 1)^n }{n!2^n (1 - s)^a (1 + s)^b}\frac{d^n }{ds^n }\left[ {(1 - s)^{a + n} (1 + s)^{b + n} }
\right],
\end{equation}

we can write

\begin{equation}
  P_n^{(a,b)} (1-2s) = \frac{C_n}{ s^a (1 - s)^b}\frac{d^n }{ds^n }\left[s^{a+n}(1-s)^{b+n}\right]
\end{equation}

and

\begin{equation}
  \frac{d^n }{ds^n }\left[s^{a+n}(1-s)^{b+n}\right]=C_n s^a (1 - s)^b P_n^{(a,b)} (1-2s).
\end{equation}

If we use the last equality in Eq.(3.28), we can write

\begin{equation}
  y_{n_{r}}(s) = C_{n_{r}} P_{n_{r}}^{(2\sqrt{c},2K-1)} (1-2s).
\end{equation}

Substituting $\phi (s)$ and $y_{n_{r}}(s)$ into Eq.(3.7), we obtain

\begin{equation}
  \chi _{n_{r}}(s)=C_{n_{r}}s^{\sqrt c}(1-s)^K P_{n_{r}}^{(2\sqrt{c},2K-1)} (1-2s).
\end{equation}

Using the following definition of the Jacobi
polynomials~\cite{Abramowitz}:

\begin{equation}
P_n^{(a,b)} (s) = \frac{{\Gamma (n + a + 1)}}{{n!\Gamma (a +
1)}}\mathop F\limits_{21} \left( { - n,a + b + n + 1,1 + a;\frac{{1
- s}}{2}} \right),
\end{equation}

we  are able to write Eq.(3.33) in terms of hypergeometric
polynomials as

\begin{equation}
  \chi_{n_{r}} (s)=C_{n_{r}}s^{\sqrt c}(1-s)^{K}\frac{\Gamma (n_{r}+2\sqrt c+1)}{n_{r}!\Gamma (2 \sqrt c+1)} \mathop F\limits_{21} \left( { - n_{r},2 \sqrt c +2K+n_{r},1 +2 \sqrt
  c;s}\right).
\end{equation}

The normalization constant $C_{n_{r}}$ can be found from
normalization condition

\begin{equation}
  \int\limits_0^\infty |R(r)|^2r^2dr=\int\limits_0^\infty |\chi (r)|^2 dr=b\int\limits_0^1\frac{1}{s}|\chi (s)|^2 ds=1,
\end{equation}

by using the following integral formula~\cite{Agboola}:

$$
 \int\limits_0^1 {(1 - z)^{2(\delta  + 1)} z^{2\lambda  - 1} } \left\{ {\mathop F\limits_{21} ( - n_{r},2(\delta  + \lambda  + 1) + n_{r},2\lambda  + 1;z)} \right\}^2 dz
$$
\begin{equation}
 = \frac{{(n_{r} + \delta  + 1)n_{r}!\Gamma (n_{r} + 2\delta  + 2)\Gamma (2\lambda )\Gamma (2\lambda  + 1)}}{{(n_{r} + \delta  + \lambda  + 1)\Gamma (n_{r} + 2\lambda  + 1)\Gamma (2(\delta  + \lambda  + 1) + n_{r})}}
\end{equation},

for $ \delta  > \frac{{ - 3}}{2}$\,\,\, and\,\,\, $\lambda >0 $.
After simple calculations, we obtain normalization constant as

\begin{equation}
  C_{n_r}=\sqrt{\frac{n_{r}!2\sqrt c(n_{r}+K+\sqrt c)\Gamma (2(K+\sqrt c)+n_{r})}{b(n_{r}+K)\Gamma (n_{r}+2\sqrt c+1)\Gamma (n_{r}+2K)} }.
\end{equation}

\section{\bf Solution of Azimuthal Angle-Dependent Part of the Schr\"{o}dinger
equation}\label{ar}

We may also derive the eigenvalues and eigenvectors of the azimuthal
angle dependent part of the  SE in Eq.(2.5) by using NU method. The
boundary condition for Eq.(2.5),  $ \Theta (\theta)$ require  to be
taken as a finite value. Introducing a new variable $x=cos\theta $,
Eq.(2.5) is brought to the form
\begin{equation}
  \Theta ''(x)-\frac{2x}{1-x^2}\Theta '(x)+\frac{1}{(1-x^2)^2}\left[\lambda (1-x^2)-m^2-(\beta^\prime +\beta x)\right]\Theta (x)=0.
\end{equation}

After the comparison of Eq.(4.1) with Eq.(3.3) we have

\begin{equation}
  \tilde{ \tau}(x)=-2x~~,\sigma (x)=1-x^2~~,\tilde{ \sigma}(x)=-\lambda x^2-\beta x+(\lambda -m^2-\beta^\prime ).
\end{equation}

In the NU method the new function $\pi (x)$ is calculated for
angle-dependent part as

\begin{equation}
  \pi (x)=\pm \sqrt {x^2(\lambda -k)+\beta x -(\lambda -\beta^\prime -m^2-k)}.
\end{equation}

The constant parameter $k$ can be determined as
\begin{equation}
  k_{1,2}=\frac{2\lambda -m^2-\beta^\prime }{2}\pm \frac{u}{2} ,
\end{equation}

where $u=\sqrt {(m^2+\beta^\prime )^2-\beta ^2}$.

The appropriate function $\pi (x)$ and parameter $k$ are
\begin{equation}
\pi (x ) =  - \left[ x \sqrt {\frac{m^2  + \beta^\prime  + u}{2}} +
\sqrt {\frac{m^2  + \beta^\prime  - u}{2}} \right],
\end{equation}

\begin{equation}
  k = \frac{2\lambda  - m^2  - \beta^\prime }{2} - \frac{u}{2}.
\end{equation}

The following track in this selection is to achieve the condition $\tau'<0$ . Therefore  $\tau(x)$ becomes

\begin{equation}
\tau (x) =  - 2x\left[ {1 +\sqrt {\frac{{m^2 + \beta^\prime  +
u}}{2}} } \right] - 2\sqrt {\frac{{m^2  + \beta^\prime -u}}{2}}.
\end{equation}

We can also write the values  $\bar\lambda =k+\pi'(s)$ as

\begin{equation}
 \bar \lambda =\frac{2\lambda -\beta^\prime-m^2}{2}-\frac{u}{2}-\sqrt{\frac{m^2+\beta^\prime +u}{2}},
\end{equation}

also using Eq.(3.20) we can equate

\begin{equation}
 \bar \lambda_{N} =\frac{2\lambda -\beta^\prime-m^2}{2}-\frac{u}{2}-\sqrt{\frac{m^2+\beta^\prime +u}{2}}=2N\left[1+\sqrt{\frac{m^2+\beta^\prime +u}{2}}\right]+N(N-1).
\end{equation}
In order to obtain unknown $\lambda $ we can solve Eq.(4.9) explicitly for $\lambda =l(l+1)$
\begin{equation}
  \lambda -\zeta ^2-\zeta =2N(1+\zeta )+N(N-1),
\end{equation}
where $\zeta =\sqrt{\frac{m^2+\beta^\prime +u}{2}},$ and
\begin{equation}
  \lambda =\zeta ^2+\zeta +2N\zeta +N(N+1)=(N+\zeta )(N+\zeta +1)=l(l+1) ,
\end{equation}
then
\begin{equation}
  l=N+\zeta .
\end{equation}

Substitution of this result in Eq.(3.23) yields the desired energy
spectrum, in terms of $n_r$ and $N$ quantum numbers. Similarly, the
wave function of azimuthal angle dependent part of SE can be
formally derived by a process to the derivation of radial part of
SE.

\begin{equation}
  \phi (x)=(1-x)^{(B+C)/2} ,
\end{equation}

\begin{equation}
\rho (x)=(1-x)^{B+C}(1+x)^{B-C},
\end{equation}

\begin{equation}
y_{N} (x) = B_N (1 -x )^{ - (B + C)} (1 + x )^{C - B} \frac{{d^N
}}{{dx^N }}\left[ {(1 - x )^{B + C + N} (1 +x )^{B - C + N} }
\right],
\end{equation}

where

$$
B=\sqrt{\frac{m^2+\beta^\prime +u}{2}}~,~C=\sqrt{\frac{m^2+
\beta^\prime -u}{2}}.
$$

From the definition of Jacobi polynomials, we can write

\begin{equation}
 \frac{d^N}{dx^N}\left[(1-x)^{B+C+N}(1+x)^{B-C+N}\right]=(-1)^N 2^N (1-x)^{B+C}(1+x)^{B-C}P_N^{(B + C,B - C)}(x).
\end{equation}
Substitution  of Eq.(4.16) into Eq.(4.15) and after long but
straightforward calculations we obtain the following result
\begin{equation}
\Theta_{N} (x) = C_N (1 - x)^{(B + C)/2} (1 + x )^{(B - C)/2}
P_N^{(B + C,B - C)} (x),
\end{equation}
where $C_N$ is the normalization constant. Using orthogonality
relation of the Jacobi polynomials ~\cite{Abramowitz} the
normalization constant can be found as
\begin{equation}
  C_N=\sqrt{\frac{(2N+2B+1)\Gamma (N+1)\Gamma (N+2B+1)}{2^{2B+1}\Gamma (N+B+C+1)\Gamma (N+B-C+1)}}.
\end{equation}

\section{\bf Numerical Results and Discussion}\label{br}

Solution of the SE for the Manning-Rosen potential plus a
ring-shaped like potential are obtained by applying  the
Nikiforov-Uvarov method in which we used the improved approximation
scheme to the centrifugal potential for arbitrary $l$ states. The
energy eigenvalues and corresponding eigenfunctions are obtained for
arbitrary $l$ quantum numbers. Two important cases must be
emphasized in the results of this study. In the first case which $
\beta =\beta ^\prime=0$ the potentials turn to central MR potential.
For this case, by using $ u=m^2$, $\zeta =| m|$ and $l=N +|
m|~(N=0,1,2...)$ then $l \geq | m| $ by substituting this $l$ values
in  Eq.(3.23) we obtain energy spectrum for MR potential. These
results are consistent with those of works in Zhao-You Chen and et
al,~\cite{Chen-Jia}. Also, if $\alpha =0 $ or $\alpha =1 $, then in
this case, for $ \delta =1/b$, $A=2b$ potential turns to the
Hulth\'en potential with $ l\geq | m| $. This energy values are
consistent with those of works in Ref.~\cite{Varshni}. The second
case is the general situation where $ \beta\neq0$ or $\beta
^\prime\neq 0$. By changing $ \beta$ and $\beta ^\prime$ we obtain
energy values of MR plus different type ring shaped like potentials.
In Table 1(for $\alpha=0.75$) and Table 2(for $\alpha=1$) we show
energies of the bound states for the Manning-Rosen potential plus
ring shaped potentials for different values of
$\beta,\beta^{'},m,l,n$ and $1/b=0.025, A=2b $, where $n=n_r+l+1$,
is the usual principle quantum number and $n_r$ is the number of
nodes of the radial wave functions. To show  accuracy of our
results, the plots of the centrifugal term $1/r^2$ (solid line), the
improved new approximation to it $\frac
{1}{r^2}=\frac{1}{b^2}[C_0+\frac{e^{-r/b}}{(1-e^{-r/b})^2}]$
(dashed) and the conventional approximation to it $\frac
{1}{r^2}=\frac{1}{b^2}[\frac{e^{-r/b}}{(1-e^{-r/b})^2}]$ (long
dashed) as function of the variable $r$ are displayed from figures 1
to 3 with different potential range parameter $b=1,2,4$. It is shown
that the our approximation Eq.(3.1) is a good approximation to the
centrifugal term for short potential range, i.e. large $b$.

Finally, we want to deal with some restrictions about bound state
solutions of SE for Manning Rosen plus ring shaped like potential.
First, it is seen from Eq.(4.4) and expression from $u$ that in
order to obtain real energy values the condition $(m^2+\beta
')^2\geq \beta ^2 $ must be hold. Since the parameters $ \beta $ and
$\beta '$ are real and positive, we can write
\begin{equation}
 m^2 \geq (\beta-\beta') .
\end{equation}

If $\beta \leq \beta'$ the inequality in Eq.(5.1) is provided
automatically. But if $\beta \geq \beta'$ then $m$ becomes bounded.
Secondly, in Eq.(3.23) if
\begin{equation}
l(l+1)C_0 > \left[
 n_{r}+1/2+\frac{(l-n_{r})(l+n_{r}+1)-A}{ 2\Lambda + 1 + 2n_{r}}
 \right]^2,
\end{equation}
 then energy eigenvalues  take  non-negative values, this means there is no bound
 states. If we take $C_0=0$ for the approaches in centrifugal term as some previous studies, the restriction on
 quantum numbers is removed. Finally, from Eq (3.21) we obtain
\begin{equation}
 \sqrt c=\frac{\lambda+1/2+\Lambda  (1 + 2n_r) + n_r(n_r + 1)-A}{ -(2\Lambda + 1 + 2n_r)} .
\end{equation}
For bound states since $ c>0$ (so $\sqrt c>0 $ ) and using the fact
that $ 2\Lambda + 1 + 2n_r>0$, we obtain
$$
 A-1/2-l(l+1)-\Lambda >
n_r(n_r+2\Lambda+1),
$$
\begin{equation}
 A> \frac{1 } {2}+l(l+1)+\Lambda .
\end{equation}
If both conditions in Eqs.(5.1-5.2 and 5.4) are satisfied
simultaneously, the bound states exist. Thus the energy spectrum
equation in Eq.(3.23) as limited, i.e, we have only the finite
numbers of energy eigenvalues. Figure 4 and 5 show the region of
possible values of $n_r$ and $l$ quantum numbers according to our
analysis in Eq.(5.2) and Eq.(5.4) for $A=80$, $\alpha=1$ and
$\alpha=0.75$, respectively.

\section{\bf Conclusion}\label{dr}

Analytical calculations of energy eigenvalues for an arbitrary $l$
state and corresponding eigenfunctions in the Manning-Rosen
potential plus a Ring-Shaped like potential is done by using
Nikiforov-Uvarov method in this paper. The energy eigenvalue
expression for Manning-Rosen potential plus a ring-shaped like
potential is given by Eq.(3.23). By some specific value of $\alpha$,
$\beta$ and $\beta ^\prime$ parameters one can observe that the some
results of previous studies for finding MR and Hulth\'en  potential
can be found. We also obtain some important restrictions on quantum
numbers about bound state solutions of SE. We can conclude that our
results are not only interesting for pure theoretical physicist but
also for experimental physicist, because the results are exact and
more general. We have also examined that, obtained results are
different from results given in ~\cite{Antia}. Therefore, the
calculation in ~\cite{Antia} should be checked and recalculated and
compared with our results and ~\cite{Chen-Jia,Varshni}.

\section*{Acknowledgments}

The author C.A thanks to member of the Department of Physics at
University of Surrey especially to Dr. Paul Stevenson giving
opportunity to research in their institute. Also thanks to Y\"{O}K
(The Council of Higher Education) for the financial support to my
research in University of Surrey.

 \newpage

\begin{table}
\begin{tabular}{|c|c|c|c|c|c|c|c|}\hline
 $\beta $ & $\beta' $& $m$  & $N$ & $n_r$ & $l$ & $n$  & $E$ \\ \hline

  0&0 &0 &0 &0 &0           &1          &-0.872300 \\ \hline
  0&0 &0 &1 &0 &0           &2          &-0.150269            \\ \hline
  0&0 &1 &0 &0 &1           &2          &-0.120527            \\
  1&1 &0 &0 &0 &1           &2          &-0.112760            \\ \hline
  0&1 &0 &0 &0 &1.414214    &2.414214   &-0.076883  \\ \hline
  1&1 &1 &0 &0 &1.618034    &2.618034   &-0.063090  \\ \hline
  0&1 &1 &0 &0 &1.732051    &2.732051   &-0.056764  \\ \hline
  0&0 &0 &2 &0 &0           &3          &-0.053926  \\ \hline
  1&0 &2 &0 &0 &1.931852    &2.931852   &-0.047497  \\ \hline
  0&0 &1 &1 &0 &1           &3          &-0.045878  \\
  1&1 &0 &1 &0 &1           &3          &-0.045878  \\ \hline
  0&0 &0 &0 &2 &2           &3          &-0.044774  \\
  1&1 &0 &0 &1 &2           &3          &-0.044774  \\ \hline
  0&1 &0 &1 &0 &1.414214    &3.414214   &-0.032286  \\ \hline
  1&1 &2 &0 &0 &2.414214    &3.414214   &-0.031721  \\
  0&1 &0 &0 &1 &2.414214    &3.414214   &-0.031721  \\ \hline
  0&1 &2 &0 &0 &2.449490    &3.449490   &-0.030831  \\ \hline
  1&1 &1 &1 &0 &1.618034    &3.618034   &-0.027392  \\ \hline
  1&1 &1 &0 &1 &2.618034    & 3.618034  &-0.026949  \\ \hline
  0&1 &1 &1 &0 &1.732051    & 3.732051  &-0.025026  \\ \hline
\end{tabular}
\qquad \qquad
\begin{tabular}{|c|c|c|c|c|c|c|c|}\hline
$\beta $ & $\beta' $& $m$  & $N$ & $n_r$ & $l$ & $n$  & $E$ \\ \hline

  0&1 &1 &0 &1 &2.732051    &3.732051   &-0.024630  \\ \hline
  0&0 &0 &3 &0 &0           &4          &-0.024017  \\ \hline
  1&0 &2 &1 &0 &1.931852    & 3.931852  &-0.021403  \\ \hline
  1& 0&2 &0 &1 &2.931852    &3.931852   &-0.021065  \\ \hline
  0&0 &1 &2 &0 &1           &4          &-0.020809  \\
  1&1 &0 &2 &0 &1           &4          &-0.020809  \\ \hline
  0&0 &0 &1 &2 &2           &4          &-0.020299  \\
  1&1 &0 &1 &1 &2           &4          &-0.020299  \\ \hline
  1&0 &3 &0 &0 &2.981188    &3.981188   &-0.020271  \\ \hline
  0&0 &1 &0 &2 &3           &4          &-0.019976  \\
  1&1 &0 &0 &2 &3           &4          &-0.019976  \\ \hline
  1&1 &3 &0 &0 &3.302776    &4.302776   &-0.015781  \\ \hline
  0&1 &3 &0 &0 &3.316625    &4.316625   &-0.015611  \\ \hline
  0&1 &0 &2 &0 &1.414214    &4.414214   &-0.015053  \\ \hline
  0&1 &0 &1 &1 &2.414214    &4.414214   &-0.014733  \\
  1&1 &2 &1 &0 &2.414214    &4.414214   &-0.014733  \\ \hline
  1&1 &2 &0 &1 &3.414214    &4.414214   &-0.014463  \\
  0&1 &0 &0 &2 &3.414214    &4.414214   &-0.014463  \\ \hline
  0&1 &2 &1 &0 &2.449490    &4.449490   &-0.014336  \\ \hline
\end{tabular}
\caption{Energies of the bound states for the Manning Rosen
potentials plus ring-shaped potentials for different values of
$\beta, \beta', m, l, n_r$ and $ \alpha=0.75, 1/b=0.025, A=2b $
calculated using Eq.(4.12) and Eq.(3.23) for $\mu=\hbar=1$.}
\label{table1}
\end{table}

\newpage
\begin{table}[h]
\begin{tabular}{|c|c|c|c|c|c|c|c|}\hline
 $\beta $ & $\beta' $& $m$  & $N$ & $n_r$ & $l$ & $n$  & $E$ \\ \hline

   0&0  &0  &0    &0   &0   &1   &-0.487578            \\ \hline
   0&0  &0  &1    &0   &0   &2   &-0.112812            \\ \hline
   0&0  &1  &0    &0   &1   &2   &-0.112760            \\
   1&1  &0  &0    &0   &1   &2   &-0.112760            \\ \hline
   0&1  &0  &0    &0   &1.414214    &2.414214   &-0.073653  \\ \hline
   1&1  &1  &0    &0   &1.618034    &2.618034   &-0.060874  \\ \hline
   0&1  &1  &0    &0   &1.732051    &2.732051   &-0.054947  \\ \hline
   1&0  &2  &0    &0   &1.931852    &2.931852   &-0.046192  \\ \hline
   0&0  &0  &2    &0   &0   &3  &-0.043759  \\ \hline
   0&0  &1  &1    &0   &1   &3  &-0.043707  \\
   1&1  &0  &1    &0   &1   &3  &-0.043707  \\ \hline
   0&0  &2  &0    &2   &2   &3  &-0.043602  \\
   1&1  &0  &0    &1   &2   &3  &-0.043602  \\ \hline
   0&1  &0  &1    &0   &1.414214    &3.414214   &-0.031215  \\ \hline
   1&1  &2  &0    &0   &2.414214    &3.414214   &-0.031089  \\
   0&1  &0  &0    &1   &2.414214    &3.414214   &-0.031089  \\ \hline
   0&1  &2  &0    &0   &2.449490    &3.449490   &-0.030230  \\ \hline
   1&1  &1  &1    &0   &1.618034    &3.618034   &-0.026609  \\ \hline
   1&1  &1  &0    &1   &2.618034    &3.618034   &-0.026473  \\ \hline
   0&1  &1  &1    &0   &1.732051    &3.732051   &-0.024363  \\ \hline
\end{tabular}
\qquad \qquad
\begin{tabular}{|c|c|c|c|c|c|c|c|}\hline
 $\beta $ & $\beta' $& $m$  & $N$ & $n_r$ & $l$ & $n$  & $E$ \\ \hline

   0&1  &1  &0    &1   &2.732051    &3.732051   &-0.024221  \\ \hline
   1&0  &2  &1    &0   &1.931852    &3.931852   &-0.020903  \\ \hline
   1&0  &2  &0    &1   &2.931852    &3.931852   &-0.020750  \\ \hline
   0&0  &0  &3    &0   &0   &4  &-0.020000  \\ \hline
   1&0  &3  &0    &0   &2.981188    &3.981188   &-0.019975  \\ \hline
   0&0  &1  &2    &0   &1   &4  &-0.019948  \\
   1&1  &0  &2    &0   &1   &4  &-0.019948  \\ \hline
   1&1  &0  &1    &1   &2   &4  &-0.019844  \\
   0&0  &1  &1    &1   &2   &4  &-0.019844  \\ \hline
   0&0  &3  &0    &0   &3   &4  &-0.019687  \\
   1&1  &0  &0    &2   &3   &4  &-0.019687  \\ \hline
   1&1  &3  &0    &0   &3.302776    &4.302776   &-0.015583  \\ \hline
   0&1  &3  &0    &0   &3.316625    &4.316625   &-0.015417  \\ \hline
   0&1  &0  &2    &0   &1.414214    &4.414214   &-0.014594  \\ \hline
   0&1  &0  &1    &1   &2.414214    &4.414214   &-0.014468  \\
   1&1  &2  &1    &0   &2.414214    &4.414214   &-0.014468  \\ \hline
   1&1  &2  &0    &1   &3.414214    &4.414214   &-0.014290  \\
   0&1  &0  &0    &2   &3.414214    &4.414214   &-0.014290  \\ \hline
   0&1  &2  &1    &0   &2.449490    &4.449490   &-0.014082  \\ \hline
\end{tabular}
\caption{Energies of the bound states for the Manning Rosen
potentials plus ring-shaped potentials for different values of
$\beta, \beta', m, l, n_r$ and $\alpha=1, 1/b=0.025, A=2b $
calculated using Eq.(4.12) and Eq.(3.23) for
$\mu=\hbar=1$.}\label{table2}
\end{table}

\newpage

\begin{figure}

\includegraphics[width=12 cm ,height=8 cm]{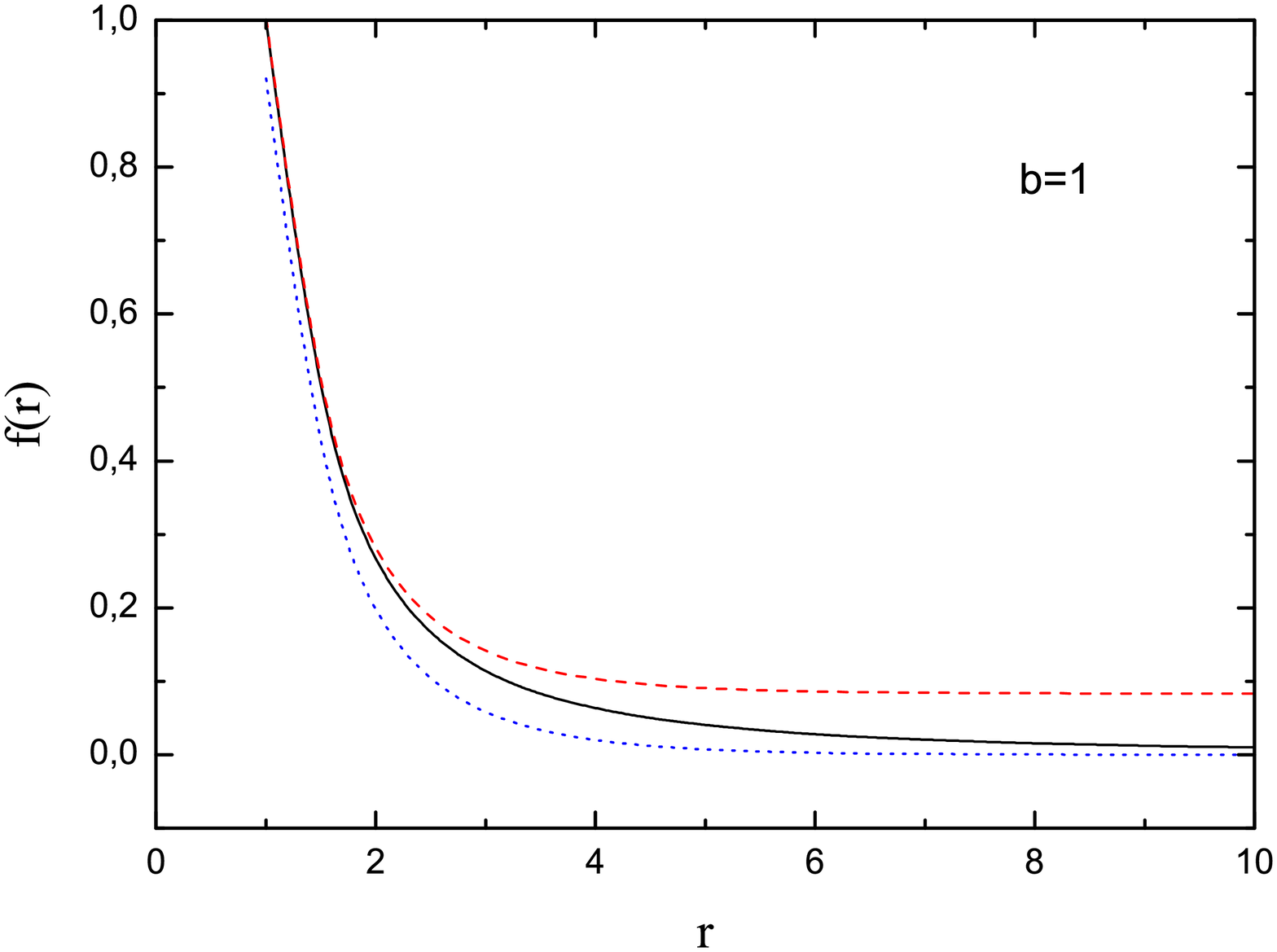}
\caption{The plots of the centrifugal term (solid line) the improved
new approximation to it (dashed) and the conventional approximation
to it (long dashed) as the function of the variable r with potential
range parameter b=1 }
\end{figure}
\begin{figure}

\includegraphics[width=12 cm ,height=8 cm]{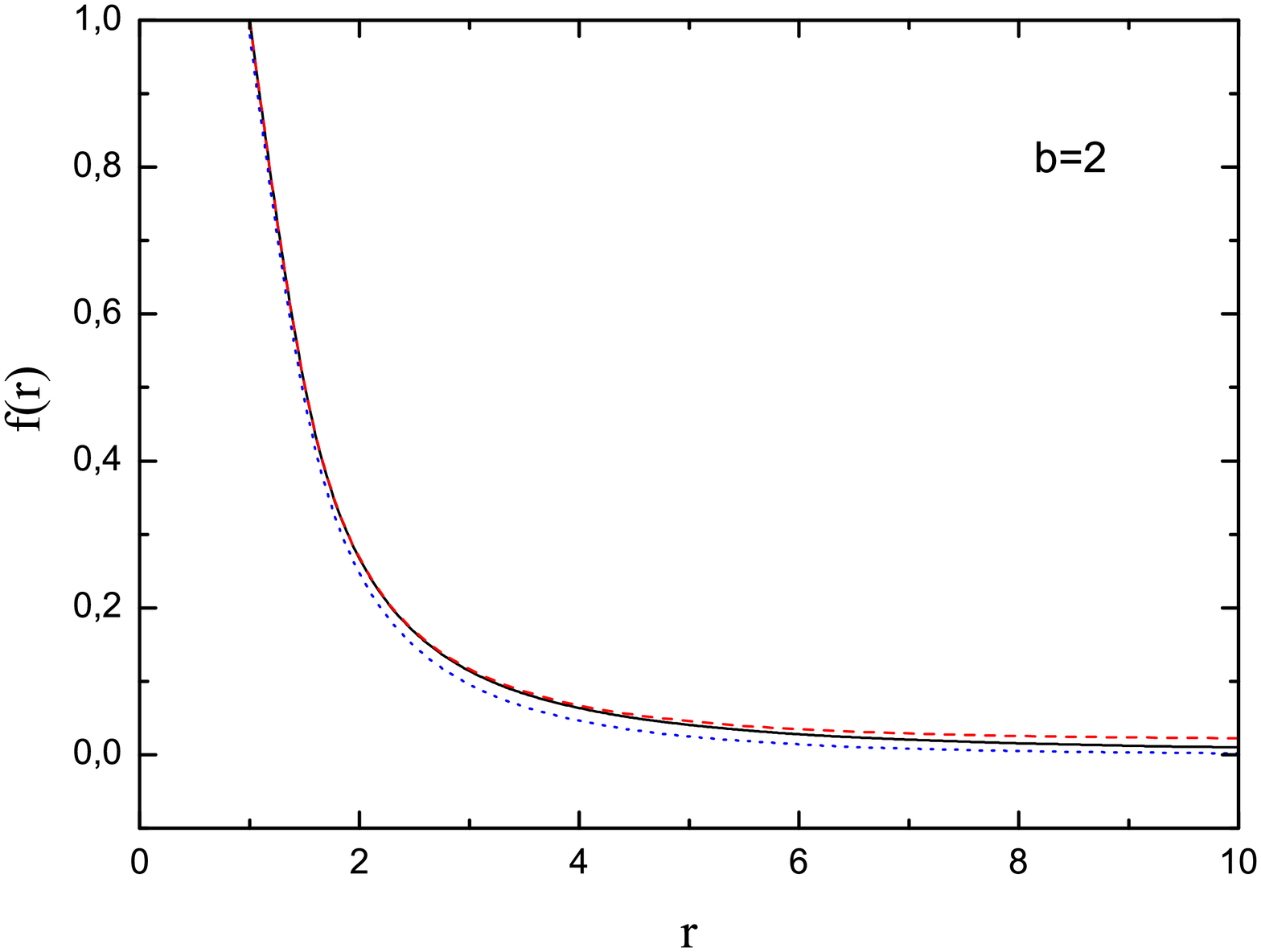}
\caption{The plots of the centrifugal term (solid line) the improved
new approximation to it (dashed) and the conventional approximation
to it (long dashed) as the function of the variable r with potential
range parameter b=2}
\end{figure}

\begin{figure}
\includegraphics[width=12 cm ,height=8 cm]{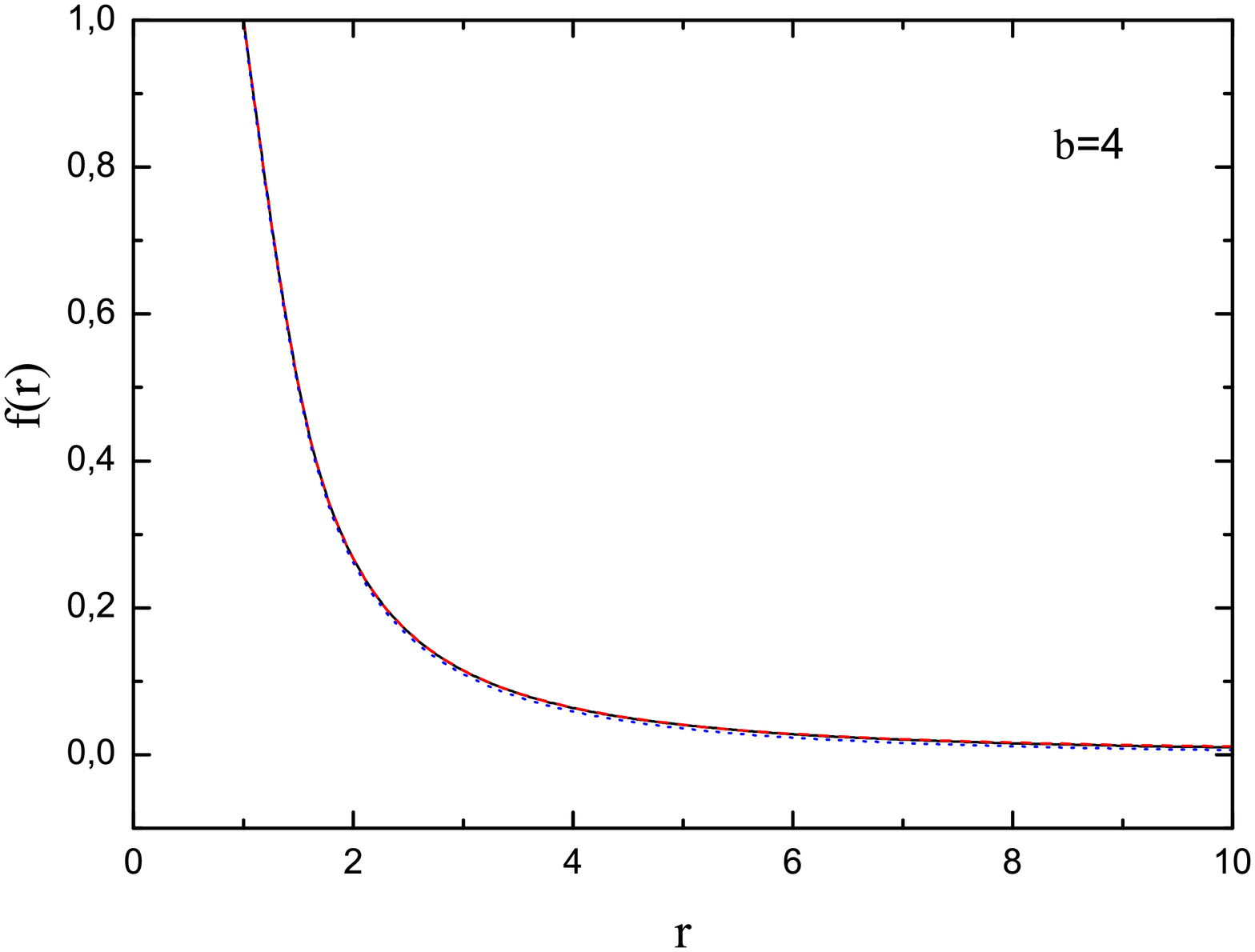}
\caption{The plots of the centrifugal term (solid line) the improved
new approximation to it (dashed) and the conventional approximation
to it (long dashed) as the function of the variable r with potential
range parameter b=4}
\end{figure}

\begin{figure}
\includegraphics[width=10 cm ,height=8 cm]{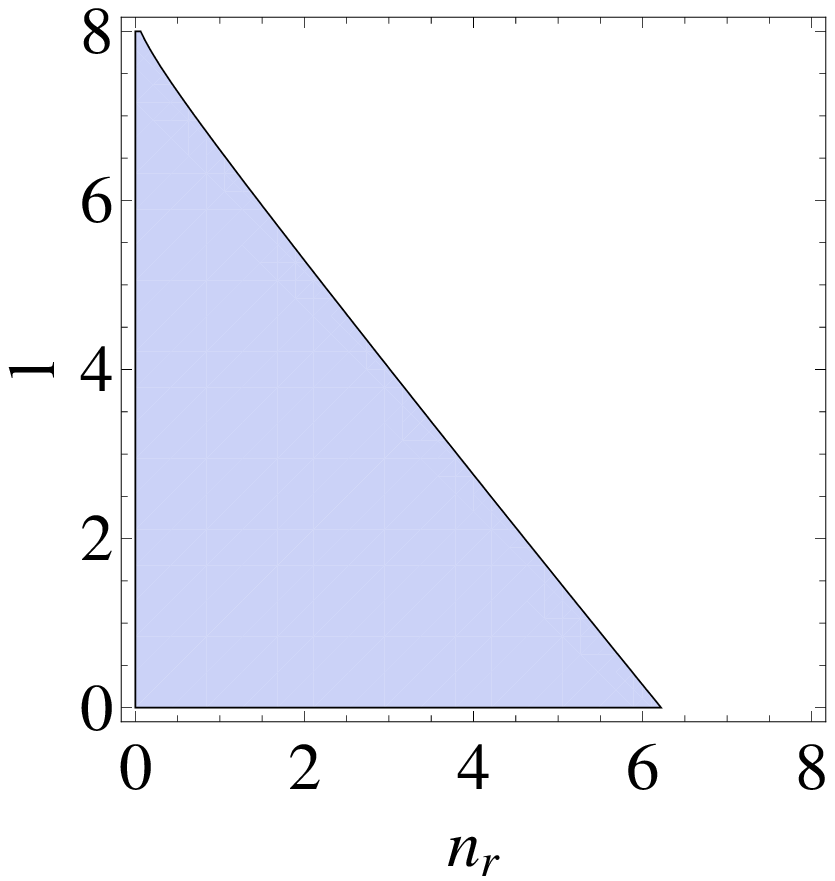}
\caption{The region of possible values of $n_r$ and $l$ for
$\alpha=0.75$, 1/b=0.025 and A=2b   }
\end{figure}

\begin{figure}
\includegraphics[width=10 cm ,height=8 cm]{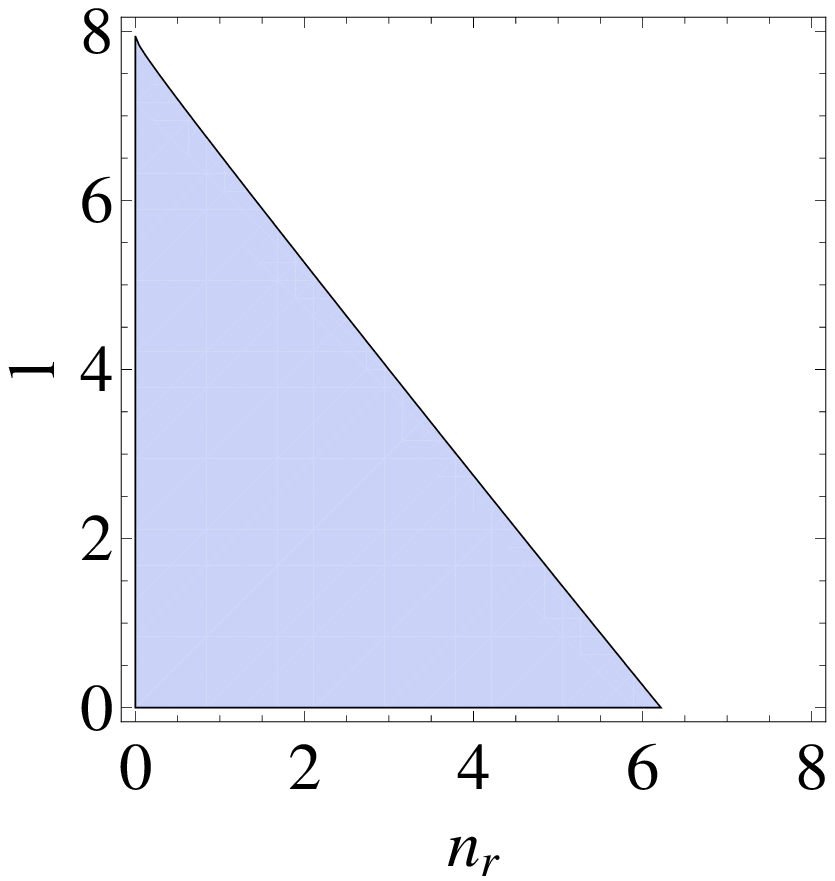}
\caption{The region of possible values of $n_r$ and $l$ for
$\alpha=1$, 1/b=0.025 and A=2b}
\end{figure}

\end{document}